\documentclass[12pt,preprint]{aastex}

\slugcomment{Submitted to {\it Astrophysical Journal Letters}}

\begin{document}

\title{SHELS: The Hectospec Lensing Survey}

\author{Margaret J. Geller\footnote{Smithsonian Astrophysical
Observatory, 60 Garden Street, Cambridge, MA 02138, USA:
e-mail: mgeller@cfa.harvard.edu, mkurtz@cfa.harvard.edu,
dfabricant@cfa.harvard.edu, ncaldwell@cfa.harvard.edu}}
\author{Ian P. Dell'Antonio\footnote{Physics Department, Brown University,
Providence, RI 02912: e-mail: ian@now.het.brown.edu}}
\author{Michael J. Kurtz$^1$}
\author{Massimo Ramella\footnote{INAF, Osservatorio Astronomico di Trieste,
via C.B. Tiepolo 11, I-34131, Trieste, Italy: e-mail: ramella@ts.astro.it }}
\author{Daniel G. Fabricant$^1$, Nelson Caldwell$^1$}
\author{J. Anthony Tyson\footnote{Physics Department,
University of California Davis, Davis, CA 95616: e-mail:
tyson@physics.ucdavis.edu; dwittman@physics.ucdavis.edu} ~ and David Wittman$^4$}

\begin{abstract}

The Smithsonian Hectospec Lensing Survey (SHELS) combines a 
10,000 galaxy deep
complete redshift survey with a weak lensing map from the Deep Lens Survey
(Wittman et al. 2002; 2005). We use maps of the velocity
dispersion based on systems
identified in the redshift survey to compare the three-dimensional
matter distribution with the two-dimensional projection mapped by
weak lensing. We demonstrate directly that the lensing map images the
three-dimensional matter distribution obtained from the kinematic data.      

\end{abstract}
\keywords {galaxies:clusters:individual (CXOU J092026+302938,  CXOU J092053+302800,
CXOU J092110+302751) --- galaxies: distances and redshifts --- gravitational lensing
--- large-scale structure of the universe}

\section{Introduction}

Weak gravitational lensing maps are a powerful
modern tool for probing the distribution of matter in the universe.  
Demonstrations of the power of weak lensing include measurements of
the masses of galaxies and galaxy groups (Dell'Antonio \& Tyson 1996; Hudson et
al. 1998;  Fischer et al. 2000;  Hoekstra et al. 2003;
Sheldon et al. 2004; Parker et al. 2005), maps of the mass distribution in
rich clusters of galaxies (Tyson et al. 1990; Squires et al. 1996; Luppino \& Kaiser 1997;
Hoekstra et al. 1998; Clowe \& Schneider 2001; Sheldon et al. 2001;
Irgens et al. 2002; Gray et al. 2002; Wittman
et al. 2003; Jee et al. 2005; Margoniner et al. 2005; Cypriano et al. 2005) 
and statistical detection of the lensing signal of the
large-scale structure of the universe (Bacon et al. 2000; Wittman et al.
2000; Wilson et al. 2001; Bernardeau et al. 2002; Hoekstra et al. 2002;
Jarvis et al. 2003; Refregier 2003; Heymans et al. 2004; van Waerbeke et
al. 2005). 
Recently wide-field imagers on large telescopes have enabled 
lensing surveys of substantial objectively chosen
regions where broader investigations of 
mass distributions in the
universe can be accomplished
(Kaiser 1998; Wilson et al. 2001; Brown et al. 2003).
Schneider (2005) reviews the impressive recent progress in this area.

The Deep Lens Survey
(Wittman et al. 2002, 2005; DLS hereafter), covers 20 square degrees in five
separate four-square-degree regions. A weak lensing (convergence) map of
one of the fields
at (09:19:32.4 +30:00:00 (J2000)) is complete (Wittman et al. 2005). 
In the DLS field, distortions of resolved objects with
$21 < R < 24.2$
reveal the matter distribution generally marked by galaxies with $ R < 21$,
the range accessible with Hectospec (Fabricant et al.
2005), a 300-fiber
moderate resolution spectrograph mounted at the $f/5$ focus of the
6.5-meter MMT.

We report initial results of combining the DLS with SHELS
(Smithsonian Hectospec Lensing Survey).
We explore
the power of coupling a
foreground redshift survey with a weak lensing map in uncovering the
three-dimensional matter distribution in
the universe from the scale of individual galaxies to
the large-scale structure itself. In this first paper we examine the
relationship between a weak lensing (convergence) map of the complete
9-hour DLS field 
with a ``velocity dispersion'' map of the same field derived from 
a\ $\sim$10,000 galaxy redshift survey to a limiting
$R < 20.3$. There is a striking correspondence between the lensing
map and the large-scale structure revealed by the redshift survey in the
redshift range $0.07 < z < 0.47$. The match between these assays of the matter
distribution shows the 
broad potential power of combining these two modern tools of astrophysics.

We begin by describing the observational approaches to the lensing map
(Section 2) and the Hectospec redshift survey (Section 3). In Section 4
we review the construction of the ``velocity dispersion map'' from the
redshift survey. 
We cross-correlate the two maps and demonstrate that
the lensing map provides a  view of the matter distribution
associated with large-scale structure in the
range covered by the Hectospec redshift survey.
We summarize  in Section 5.

\section{The DLS: The Deep Lens Survey Weak Lensing Map}

The DLS (Wittman et al. 2002; 2005) is an NOAO Survey Project using the
Mosaic I and II cameras on the 4m telescopes on Kitt Peak and Cerro Tololo. 
The effective exposure time of the survey is about 14500 seconds in
R and 
the 1 $\sigma$ surface brightness limit in R is 28.7
magnitudes per square arcsecond, yielding about 45 resolved sources per
square arcminute.

The DLS  observing procedures described in Wittman et al. (2002, 2005)
produce 40$^{\prime}\times
40^{\prime}$ subfields of uniform image quality. 
To optimize the
number of sources usable for construction of the weak lensing map,
the R-band observations
are all made in seeing better than 0.9$^{\prime\prime}$.

    \begin{figure}
     \plottwo{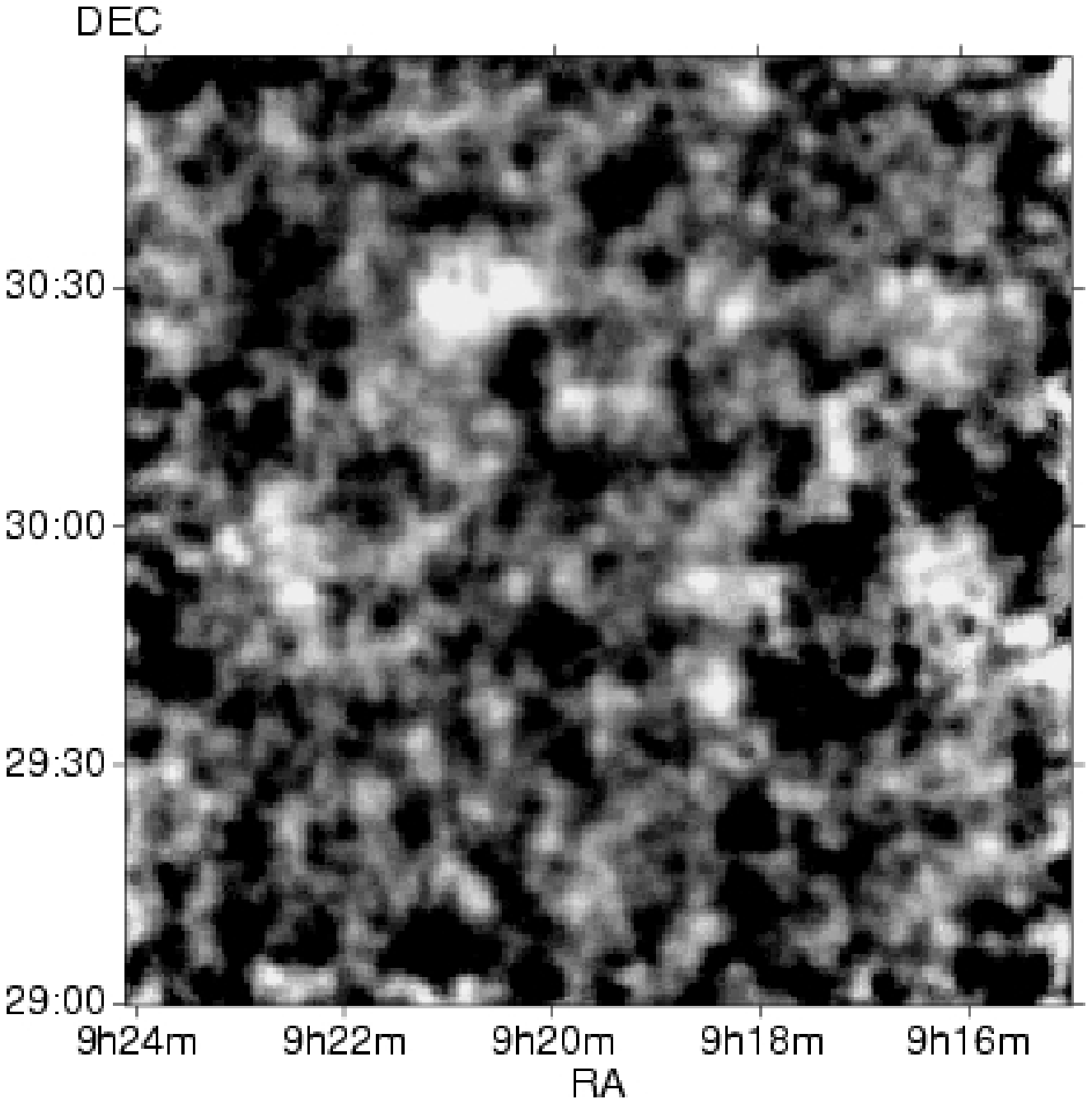}{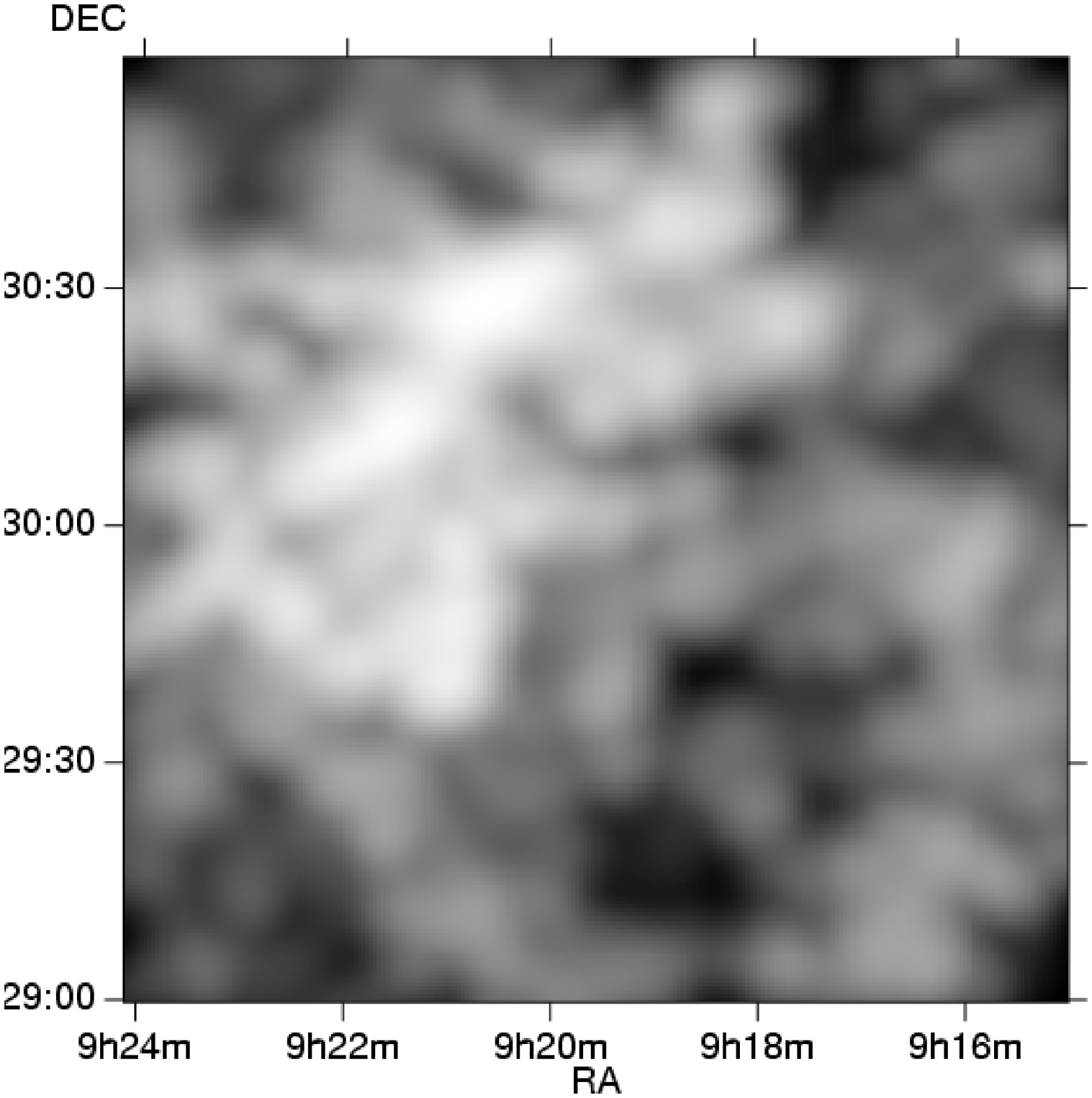}
     \caption{Convergence map for the DLS field centered at
09:19:32.4 +30:00:00 (left). The $\sigma^2$ map covering the redshift
range 0.13-0.47 (right). The range of $\kappa$ is 0 --- 0.024;
the range of $\sigma$ is $0 - {800}$ km s$^{-1}$. In both images 0
is black; the peak value is white.
}
      \end{figure}

The convergence map we construct 
for the DLS field centered at 09:19:32.4 +30:00:00 (Figure 1a) is
somewhat shallower than the
map of this field discussed by Wittman et al. (2005).  Our
convergence map is based on all resolved objects
(size at least 1.3 times the local PSF) with R band aperture
magnitude (within a 5$^{\prime\prime}$ diameter aperture) between 21
and 24.2, 
smaller than 8\arcsec\ (to exclude nearby LSB galaxies), and 
well-measured (c.f. Bernstein \& Jarvis 2002).
Approximately $2\times 10^5$ galaxies contribute to the map.
We used a Hectospec deep field
covering the range 21 $ < {\rm R} < 22.5$ (Fabricant et al. 2005)
to estimate the  weighted mean redshift of the sources 
which lies between 0.60 and 0.85.

The convergence map is a  direct reconstruction (Kaiser \&
Squires 1993) 
following the procedure of Fischer \& Tyson (1997). We convert the measured shear
into a measurement of the convergence at position ${\bf x}$ with a weight function introduced to
prevent noise divergences:

$$\kappa({\bf x}) = {\Sigma({\bf x}) \over \Sigma_{crit}}  = {1\over
N\pi}\sum_{g}
 {W(|{\bf x_g-x}|) {{e_T\bf( x_g-x)}\over |{\bf x_g-x}|^2}}$$
where $e_T\bf( x_g-x) $ is the projection of the (seeing-corrected) 
ellipticity tangent to the vector connecting ${\bf x}$ and ${\bf x_g}$, the
position of source ${\bf g}$,  and 
$W(x) = (1-e^{-x^2/(2r_i^2)})e^{-x^2/(2r_o^2)}$ and $r_i$
and $r_o$ are inner and outer cutoffs to the summation.
We used $r_i=3$\arcmin\ and $r_o=30$\arcmin.
The choice of inner cutoff radii and the source density determine the
final resolution of the map, $\sim$ 2.5\arcmin\

\section{SHELS: The Hectospec Redshift Survey}

Our goal is to understand the relationship between the 
convergence map (Figure
1a) and the distribution of matter marked
by foreground galaxies. As a first step toward that
goal we used the Hectospec (Fabricant et al. 1998; Fabricant et al. 2005),
a 300-fiber moderate resolution spectrograph mounted at the f/5 focus of the MMT
to carry out a redshift survey of the region.

We constructed a galaxy catalog from the R-band source list for the 
field. We separated galaxies from stars by central surface brightness and
then selected
galaxies with $R < 20.3$. The Hectospec 
positioning software (Roll et al. 1998) enables
optimization of the target list to obtain
an essentially complete magnitude limited survey. 
Our 46  Hectospec
pointings in the DLS field yield
reliable redshifts for 9792 galaxies with a
median redshift $ z = 0.297$.

The Hectospec spectra cover the wavelength range $\lambda 3500-10,000$
\AA \ with a resolution
of$\sim 6$ \AA. The Hectospec exposures
times are 0.75 - 1.0 hours. We reduce the data using the standard
Hectospec pipeline
(Mink et al. 2005). We derive
redshifts by application
of RVSAO (Kurtz \& Mink 1998) with templates constructed for
this purpose (Fabricant et al. 2005).
Repeat observations of 343 galaxies
imply a mean external error in the redshifts of 45 km s$^{-1}$  for
absorption-line galaxies and 30 km s$^{-1}$ for emission line galaxies.

The  redshift survey is 95\% complete for $R < 19.7$. The differential
completeness is 65\% in the range R = 19.7-20.3. Throughout the
apparent magnitude range, the redshift survey
completeness is uniform over the entire DLS field.
 
\section{A Lensing Map ... A Redshift Survey}

Here we examine the correspondence
between the large-scale structure revealed by the redshift survey and
the structure in the lensing map. The lensing map shows a complex of
three overlapping peaks, DLSCL J0920.1+3029, corresponding to three extended cluster
x-ray sources (Wittman et al. 2005). The Hectospec redshift
survey identifies three clusters coincident with the x-ray clusters and
with the convergence map peaks. CXOU J092026+302938 (Abell 781, A hereafter) has a mean
redshift 0.302 and a rest frame line-of-sight velocity dispersion
$\sigma_{A}$ = 674$^{+43}_{-52}$ km s$^{-1}$ (163 members).
CXOU J092053+302800 (B hereafter) has a mean
redshift 0.291 and $\sigma_B$ = 741$^{+35}_{-40}$ km
s$^{-1}$ (123 members).  
Figure 2 shows the complex structure in the redshift survey
in the range of these two
systems; they are separate
clusters, not subclumps of a single system. 
CXOU J092110+302751, although very close to the other two
clusters on the sky,
has a mean redshift of 0.427 and a velocity dispersion
of 733$^{+77}_{-112}$ km s$^{-1}$ (33 members). 

    \begin{figure}
     \plotone{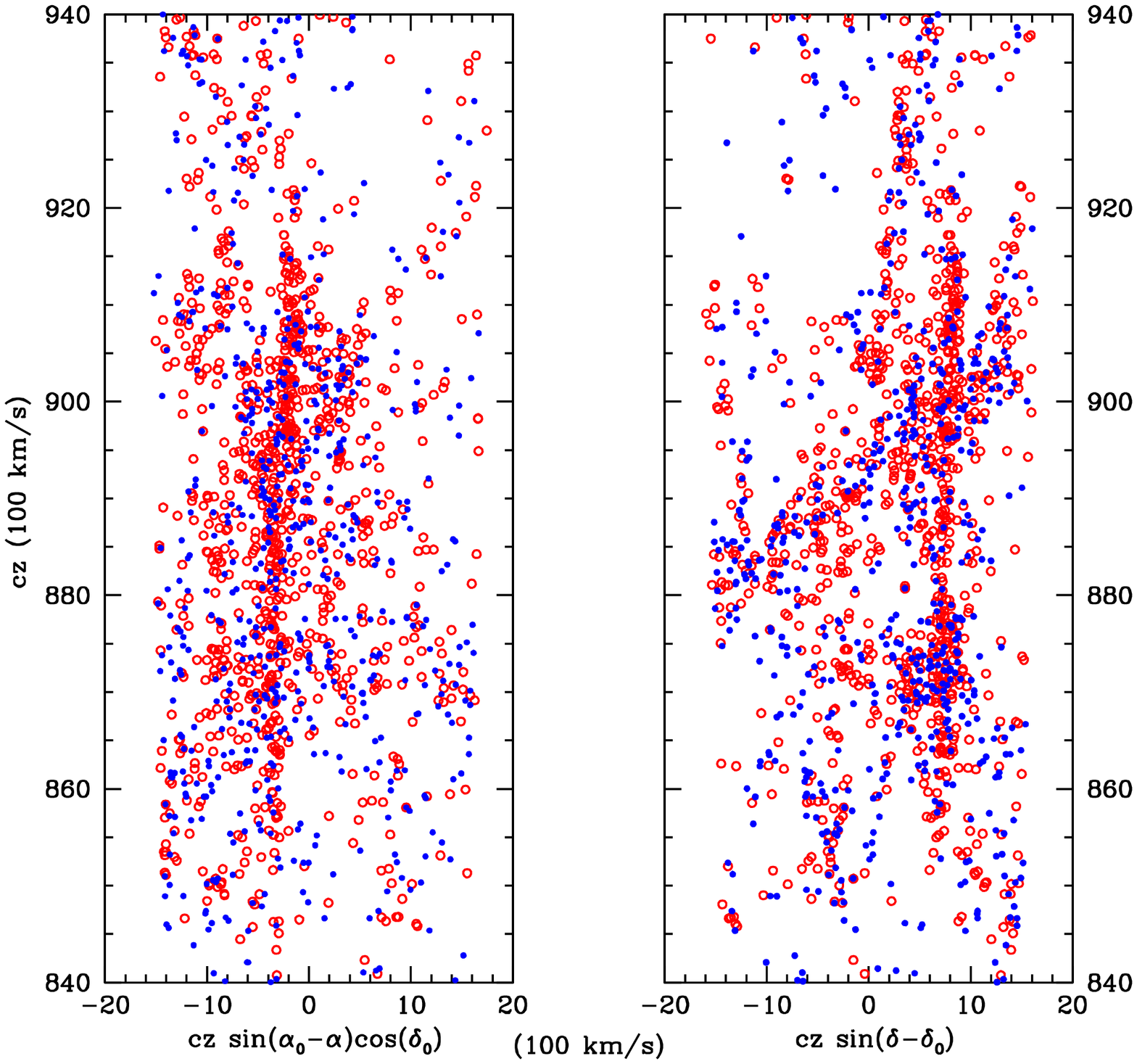}
     \caption{Section of the redshift survey including A781. Open
(red) circles denote absorption-line galaxies; filled (blue) circles denote
emission-line galaxies. There are 1590 galaxies in the plot. The left
panel is a projection of the  full declination range in right ascension;
the right is a projection in declination.}
      \end{figure}

Clusters A and B are the two highest amplitude peaks in the convergence
map. The ratio of convergence
peak amplitudes, ${p_A / p_B} = 1.02 \pm 0.10$, is consistent
with the ratio of ${\sigma_A^2 / \sigma_B^2} = 0.83 \pm 0.07$.
In this comparison, we do not attempt to account for the superposition of the outer regions
of the cluster mass distributions.

In addition to these clusters, the lensing map shows
structure at lower amplitude throughout the field. We might expect to see corresponding
foreground large-scale structure marked by groups of galaxies
(Vale \& White 2003). To identify these potentially
corresponding foreground condensations,
we apply a friends-of friends groups finding algorithm to the redshift
survey. The algorithm we use scales the
linking length to account for the variation in the observed density of a 
magnitude limited survey as a function of redshift (Ramella, Pisani \&
Geller 1997). The maps we construct are stable for
density contrasts in the range 300-1200; we use maps
from two catalogs of
condensations with number density contrast 400 and 600
in the redshift range 0.07-0.47.
The fiducial
line-of-sight linking length is 180 km s$^{-1}$.
At these
substantial contrasts the
results are robust and they are insensitive to incompleteness in the 
survey in the sense that we may miss real condensations but the ones we
identify provide a stable estimate of the local velocity dispersion.
The condensations lying closest to
peaks in the lensing map all have at least 5 members.

The groups we identify in the Hectospec redshift survey range in velocity dispersion
from$\sim{100}$ km s$^{-1}$ to $\sim$800 km s$^{-1}$. These systems
contain $\sim{40}$\% of the galaxies in the redshift survey. 
We use the ensemble of members of these systems to construct an approximate
projected mass density map to compare with the lensing map. We divide the
DLS field into a uniform 100 $\times$ 100 1.2$^{\prime}$ grid. At each density contrast we 
run through the list of group members and assign the 
square of the velocity dispersion, $\sigma^2$, of
its parent group to the nearest node of the grid. If members of different
groups are assigned to the same node, we assign the maximum group $\sigma^2$
to the node.
We then have two lists (planes) of nodes and tags, one for each 
of the two number density contrasts.
We sum the two planes and smooth the resultant map
with a 2 pixel (2.4$^{\prime}$) Gaussian.

Figure 1b shows the $\sigma^2$ map covering the redshift
range 0.13-0.47.   
The appearance of Figure 1b is
insensitive to reasonable variations in the
linking parameters, the group finding algorithm, or details of the construction method. 
The similarity between Figure 1b and the
lensing map in Figure 1a is striking even discounting the
A781 complex. There are some differences in both directions. There are
significant lensing peaks which have no counterpart in the redshift survey.
Within the sensitive range of the lensing
survey, there are also dense, well-populated systems in the redshift survey
which have no counterpart in the lensing map.
Ramella et al. (2005, in preparation) report
on the detailed match of individual systems and convergence peaks.
\begin{figure}
     \plotone{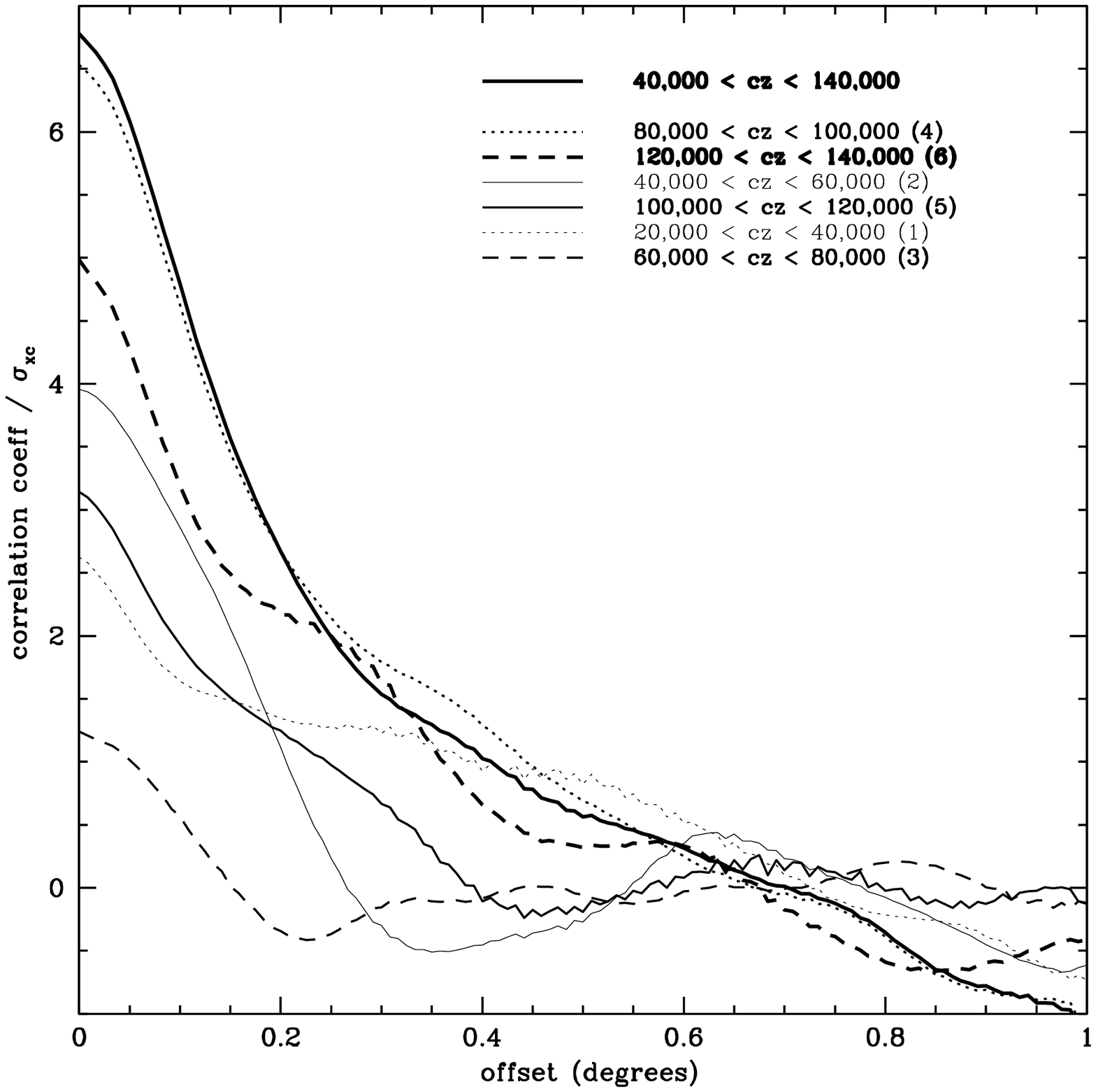}
     \caption{Cross-correlation of $\sigma^2$ maps with the convergence
map (Figure 1a). The figure inidicates the $cz$ range for each $\sigma^2$
map.}
      \end{figure}

Figure 3 shows the azimuthally averaged cross-correlation of
the convergence and $\sigma^2$  maps
in Figures 1a and 1b, respectively.
We perform the image correlations by wrapping the reference image, 
top to bottom and left to right, equivalent to the surface of a torus; 
this approach preserves the proper off-peak noise behavior.   
The cross correlation is significant at the 6.7$\sigma_{XC}$ level
where $\sigma_{XC}$  is the noise in the cross-correlation at zero lag. We
evaluate  $\sigma_{XC}$
by cross-correlating the $\sigma^2$ map with 170 noise maps.
We construct noise maps
by randomly assigning the j-th galaxy's shape and
orientation to the i-th galaxy coordinate.
This procedure preserves the spatial sampling
of the lensing map but wipes out the expected tangential ellipticity signal,
and should result in a map with the same noise
properties as the original map but none of the signal.
The mean cross-correlation between any $\sigma^2$ map and 
the noise maps is consistent with zero; the variance in the
noise map cross-correlation at zero lag,
essentially the same for all of the $\sigma^2$
maps, is our estimate of $\sigma_{XC}$.

Figure 3 also shows the azimuthally averaged cross-correlation between the
convergence 
and $\sigma^2$ maps of 6
slices each covering the redshift range $\Delta (z) = 0.067$.
Taken together the 6 slices cover
$z = 0.06 - 0.47$ (they are numbered sequentially (1)-(6) from low to high
$z$ in Figure
3). We construct
these redshift slice maps with the same method
used for the map in Figure 1b. 
Two rich clusters (Figure 2) lie in slice (4) which returns the largest
amplitude cross-correlation among the 6 slices. The clusters
and several rich groups associated with subsidiary peaks in the lensing map
lie in an extensive sheet-like structure which runs diagonally across the
region and contributes much of the high amplitude structure in Figure 1b. 

The cluster CXOU J092110+302751 at $z = 0.427$
lies within  slice (6) which has a cross-correlation amplitude of 4.9
$\sigma_{XC}$. 
The cross-correlation signal may be enhanced by superposition: the
outer regions of clusters in slice (4) overlap CXOU J092110+302751 and
groups surrounding  the cluster are superposed on systems in
the lower redshift slice (4).

Slices (2) and (5) contain no systems with $\sigma >500$ km s$^{-1}$
and yet the cross-correlation is significant. The signal comes from the
large-scale structure marked by groups of galaxies;
in both slices the several  systems which
contribute the most signal have velocity dispersions in the range 400-500 km
s$^{-1}$. For slice (2) the structure in both the convergence and sigma maps
is in the area 9:15:10.1 $ < \alpha_{2000} <$ 9:17:04.5
and 29:00:53 $ < \delta_{2000} <$ 30:01:26 ; for slice (5) it is
in the range 9:14:50.9 $ < \alpha_{2000} <$ 9:18:22.7
and   29:44:25 $ < \delta_{2000} <$ 30:43:09. These slices
demonstrate that the convergence map images the large-scale structure
marked by groups.

Slice (3) provides further insight into the relationship between the
foreground large-scale structure and the convergence map. The redshift range
of slice (3) contains several large voids. Most of the groups in this
redshift range have low velocity dispersions; the slice contains only one
system with a velocity dispersion of $\sim$400 km s$^{-1}$
along with several in the 250-300 km s$^{-1}$ range.
The cross-correlation amplitude is not significant as might be expected from
the nature of the large-scale structure in the slice.

Slice (1) with a median $z = 0.1$ provides an interesting test of
the impact of lensing efficiency  on the cross-correlation signal. The
slice contains a cluster at $z\sim{0.12}$ with a velocity dispersion of 
$\sim{740}$ km s$^{-1}$, very similar to our x-ray clusters at $z = 0.3-0.4$.
The amplitude of the cross-correlation is only $ 2.6 \sigma_{XC}$.
The ratio of lensing efficiencies for similar systems at $z = 0.1$ and
$z = 0.3$ is ${0.65 \pm 0.06}$ for the DLS sources, probably accounting for the
low cross-correlation amplitude.

Figure 3 is  a direct demonstration that
weak lensing images the foreground large-scale structure throughout the
redshift range 0.13 --- 0.47. The lensing signal is strongly modulated by
the details of the three-dimensional matter distribution. At
redshifts $ < 0.13$ we
observe a decline in lensing effectiveness. 

\section{Conclusions}

SHELS combines the power of a deep complete 10,000-galaxy redshift survey
with weak lensing maps from the DLS to examine the three-dimensional matter
distribution in the universe over the redshift range 0.07-0.47. We use
$\sigma^2$ maps based on the 
objective identification of groups and clusters in the redshift survey as a
proxy for the matter distribution in narrow redshift slices.
Cross-correlation of these $\sigma^2$ maps with the DLS weak lensing map
demonstrates that weak lensing images the foreground large-scale structure
marked by groups with velocity dispersions as low as 400 km s$^{-1}$.
The lensing signal is modulated by the large-scale structure and by the 
expected lensing efficiency.

We are extending SHELS to  greater redshift in the
region discussed here. We are also carrying out a similarly dense 
and deep redshift survey of a second DLS field.
We soon plan to report in detail on further insights
gained by combining deep dense
redshift surveys with weak lensing convergence maps.

\acknowledgments 
We thank P. Berlind and M. Calkins for their expert
operation of the Hectospec; D. Mink, J. Roll, S. Tokarz, and W. Wyatt for
constructing and running of the Hectospec pipeline;
S.Kenyon and M. Zaldarriaga for many discussions. The Smithsonian Institution
generously supported Hectospec and SHELS. Lucent Technologies and
NSF grants AST 04-41-72 and AST
01-34753 generously supported the DLS.  We used  IRAF, ADS, 
WCSTools, and SAOImage DS9 developed
by the Smithsonian Astrophysical Observatory. We appreciate generous
allocations of telescope time on the KPNO 4-meter and on the MMT.

{\it Facilities:} \facility{MMT (Hectospec)}, \facility {Mayall
(MOSAIC-I and II wide-field cameras)}

{}
 \end{document}